\documentclass[nofootinbib,aps,twocolumn,preprintnumbers]{revtex4}

\usepackage{enumitem}
\usepackage[referable]{threeparttablex}
\renewlist{tablenotes}{enumerate}{1}
\makeatletter
\setlist[tablenotes]{label=\tnote{\alph*},ref=\alph*,itemsep=\z@,topsep=\z@skip,partopsep=\z@skip,parsep=\z@,itemindent=\z@,labelindent=\tabcolsep,labelsep=.2em,leftmargin=*,align=left,before={\footnotesize}}
\makeatother

\usepackage{graphicx}  
\usepackage{subfigure}
\usepackage{multirow}
\usepackage{mathtools}
\usepackage{color}
\usepackage{ulem}
\usepackage{array}
\usepackage{diagbox}
\usepackage{comment}
\usepackage[version-1-compatibility]{siunitx}
\usepackage{siunitx}
 \usepackage{url}

\usepackage{fancyhdr}
\usepackage{longtable}
\usepackage{parskip}
\usepackage[T1]{fontenc}
\usepackage{dcolumn}   

\usepackage{bm}        
\usepackage{amsfonts}  
\usepackage{amsmath}   
\usepackage{amssymb}   

\newcommand{\pwisein}{\left\{ \begin{array}{ll}}
\newcommand{\pwiseout}{\end{array}\right.}

\begin{document}

\title{Protocol for Optically Pumping AlH$^+$ to a Pure Quantum State}

\author{Panpan Huang}
\affiliation{Department of Physics and Astronomy, Northwestern University, Evanston, IL 60208, USA}

\author{Schuyler Kain}
\affiliation{Department of Physics and Astronomy, Northwestern University, Evanston, IL 60208, USA}

\author{Antonio de Oliveira-Filho}
\affiliation{Departamento de Qu\'{i}mica, Laborat\'orio Computacional de Espectroscopia e Cin\'etica, Faculdade de Filosofia, Ci\^encias e Letras de Ribeir\~ao Preto, Universidade de S\~ao Paulo, Ribeir\~ao Preto-SP 14040-901, Brazil}

\author{Brian C. Odom}
\affiliation{Department of Physics and Astronomy, Northwestern University, Evanston, IL 60208, USA}

\date{\today}

\begin{abstract}  

We propose an optical pumping scheme to prepare trapped $\mathrm{AlH}^+$ molecules in a pure state, the stretched hyperfine state $\lvert F=\frac{7}{2},\, m_F=\frac{7}{2}\rangle$ of the rovibronic ground manifold $\lvert \mathrm{X}^2\Sigma^+,\, v=0,\, N=0\rangle$. Our scheme utilizes linearly-polarized and circularly-polarized fields of a broadband pulsed laser to cool the rotational degree of freedom and drive the population to the hyperfine state, respectively. We simulate the population dynamics by solving a representative system of rate equations that accounts for the laser fields, blackbody radiation, and spontaneous emission. In order to model the hyperfine structure, new hyperfine constants of the $\mathrm{A}^2\Pi$ excited state were computed using a RASSCF wavefunction. We find that adding an infrared laser to drive the $1$\,--\;$0$ vibrational transition within the $ \mathrm{X}^2\Sigma^+$ manifold accelerates the cooling process. The results show that under optimum conditions, the population in the target state of the rovibronic ground manifold can reach 63 $\%$ after 68 \si{\micro}s (330 ms) and 95 $\%$ after 25 ms (1.2 s) with (without) the infrared laser.

\end{abstract}


\maketitle 

\begin{figure*}[!htp]
\centering
\includegraphics[width=18cm]{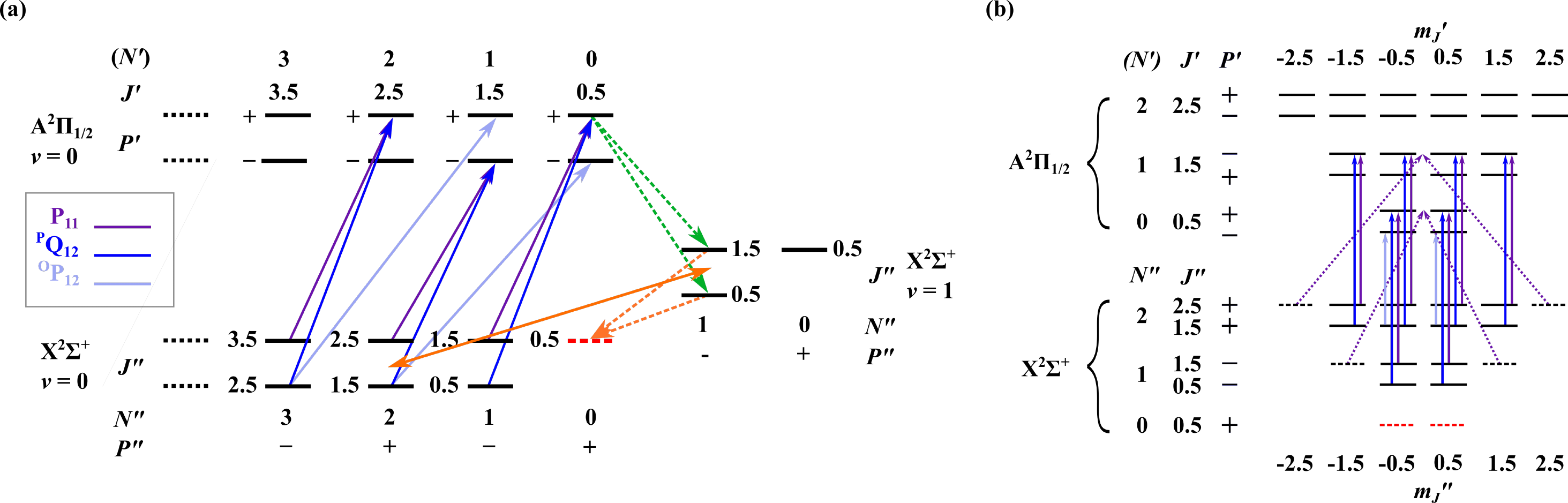}
\caption{(a) Rotational cooling mechanisms of $\mathrm{AlH}^+$ and (b) the energy level structure of $\mathrm{AlH}^+$. In (a), the rotational cooling laser that drives the $\mathrm{P_{11}}$, $\mathrm{^PQ_{12}}$ and $\mathrm{^OP_{12}}$ branches from $\lvert \mathrm{X}^2\Sigma^+, v=0\rangle$ to $\lvert \mathrm{A}^2\Pi_{1/2}, v=0\rangle$ are shown. Though the rotational angular momentum is not a good quantum number in $\mathrm{A}^{2}\Pi$, $N'$ is used here as a convenient label for different $J$ values. The dashed green arrows represent the electronic spontaneous emission with vibrational excitation. The dashed orange arrows represent the rovibrational relaxation within the $\mathrm{X}$ state. The solid orange arrow represents the rovibrational transition driven by the V10 laser. The dashed red line is the rovibronic ground state. In (b), the states are arranged vertically to reflect their approximate energies; the colors of the arrows represent the relative transition energies. States that would be dark to the cooling laser if it were not for the transverse component of the magnetic field are represented by dashed black lines. The dotted arrows do not connect any particular final m$_J$-states; they indicate that the initial dark-state populations are transferred to a mixture of m$_J$ states, thus returning them to the cooling cycle. The target states of rotational cooling, namely the rovibronic ground states, are represented by dashed red lines.}
\label{level_diagram_parity}
\end{figure*}

\section{Introduction}
\ifnum 1=0
In recent years, molecules have played increasingly important roles 
in many active areas of AMO research such as precision 
measurement\cite{safronova2018search, andreev2018improved, 
alighanbari2020precise, chou2020frequency}, quantum 
simulation\cite{ohmori2017special}, quantum information 
processing\cite{ni2018dipolar, hudson2018dipolar, 
campbell2019dipole} and cold chemistry\cite{balakrishnan2016perspective, 
bohn2017cold}. For example, new physics may be revealed through the variations 
in the proton-to-electron mass ratio. The change to the transition frequency 
caused by the variation can be enhanced utilizing the natural sensitivity 
differences between the internal states of molecules.\cite{demille2008enhanced, kobayashi2019measurement, kokish2018prospects, stollenwerk2018optical}. Modern measurements of the electric dipole moment (EDM) of the electron exploit the internal electric fields of polar molecules\cite{andreev2018improved}. 
The rich internal degrees of freedom and the long-range dipole-dipole interaction between polar molecules offer possibilities of developing a quantum toolkit for both quantum simulation and quantum information processing\cite{wei2011entanglement}. A chemical reaction may be controlled by suitable preparation of the external and internal states of the reactants\cite{de2011controlling}.

Many of these applications typically study isolated ensembles of molecules and employ various traps to achieve long interrogation times. Sometimes the measurement or interaction process is irreversible and new ensembles must be quickly prepared in the same initial state to achieve precision or statistical significance. Ion traps are commonly employed because their deep potential depths suppress the loss of trap population due to collisions with background gas. Trapped molecular ions are often sympathetically cooled with co-trapped Doppler-cooled atomic ions, leading to longer trapping lifetimes and making them well-suited for applications requiring long interrogation time. Such long hold-times also enable single-molecule studies and permit quantum states to be reinitialized \textit{in situ}.\par
\fi

\ifnum 1=0
In recent years, molecules have played increasingly important roles in many active areas of AMO research such as precision measurement\cite{safronova2018search, andreev2018improved, 
alighanbari2020precise, chou2020frequency}, coherent cold chemistry\cite{balakrishnan2016perspective, 
bohn2017cold}, quantum information processing\cite{ni2018dipolar, hudson2018dipolar, 
campbell2019dipole}, and quantum simulation\cite{ohmori2017special}. For example, new physics may be revealed through a variation in the proton-to-electron mass ratio. The change to the transition frequency caused by the variation is enhanced by the natural sensitivity differences between the internal states of molecules\cite{demille2008enhanced, kobayashi2019measurement, kokish2018prospects, stollenwerk2018optical}. Modern measurements of the electric dipole moment (EDM) of the electron exploit the internal electric fields of polar molecules\cite{andreev2018improved}. Chemical reactions may be controlled by suitable preparation of external and internal states of reactants\cite{de2011controlling}. The rich internal degrees of freedom and the long-range dipole-dipole interaction between polar molecules offer possibilities of developing a quantum toolkit for quantum information processing and simulation\cite{wei2011entanglement}.

Central to the last two applications is the development of qubits. Qubits can be represented by the pure states of atoms and molecules. However, molecules are typically more polarizable than atoms. This makes molecules more amenable for use in quantum logic gates that are addressed and manipulated by light and fields. While quantum logic gates have already been prototyped, ideal systems with pure states that are easy to define and have long coherence times remain an intense topic of investigation.

Other important requirements of quantum computers and simulations include scalability and long interrogation times. Sometimes a measurement or interaction process is irreversible and new ensembles must be quickly prepared in the same initial state to achieve readout precision. Studies of isolated ensembles of molecules typically employ various traps to achieve long interrogation times. In particular, ion traps are attractive because their deep potential depths suppress the loss of trap population due to collision with background gas. Trapped molecular ions are often sympathetically cooled with co-trapped Doppler-cooled atomic ions, leading to longer trapping lifetimes and making them well-suited for applications requiring a long interrogation time.
\fi
In recent years, molecules have played increasingly important roles in many active areas research such as precision measurement\cite{safronova2018search, andreev2018improved, 
alighanbari2020precise, chou2020frequency} and cold chemistry\cite{balakrishnan2016perspective, bohn2017cold, de2011controlling}. Also, the rich internal degrees of freedom and long-range dipole-dipole interaction between polar molecules offer possibilities of developing a toolkit for quantum information processing and quantum simulation\cite{ohmori2017special, wei2011entanglement, ni2018dipolar, hudson2018dipolar, campbell2019dipole}. Central to these applications is the development of qubits, which can be represented by the pure states of atoms or molecules\cite{chou2017preparation}. The long hold times, environmental isolation, and quantum control demonstrated for atomic ions in radiofrequency traps have made these platforms popular for work on atomic qubits. As for atomic qubits, one critical feature of molecular qubits is the ability to rapidly reset them into pure quantum states. In this work, we explore the use of optical pumping to prepare trapped molecular ions in pure states.

Our group has previously shown that rotational cooling of diatomic molecules can be achieved using a spectrally-filtered femtosecond laser (SFFL) with species that have relatively large rotational constants and fairly diagonal Frank-Condon factors (FCFs)\cite{lien2011optical}. One such example is aluminum monohydride cation, $\mathrm{AlH}^+$, for which we demonstrated an increase in the rotational ground state population from a few percent to $\sim$ 95 $\%$ within a second \cite{lien2014broadband}. The cooling of the ions to a single rotational Zeeman state was also theoretically investigated using the approach of optimal control theory \cite{aroch2018optimizing}. However, the operation of cooling $\mathrm{AlH}^+$ to a specific hyperfine state has not yet been addressed. Such hyperfine cooling has been demonstrated on the molecular ion, HD$^+$, with the transfer taking a few tens of seconds and the target population reaching 19 \%\cite{bressel2012manipulation, schneider2010all}. Also, a quantum-logic technique has demonstrated the ability to project a single trapped molecular ion into a pure state\cite{chou2017preparation}.

This manuscript proposes an efficient method to transfer the $\mathrm{AlH}^+$ population to a single stretched hyperfine state of the rovibronic ground manifold and is organized as follows. In Section II, we review the theory and describe our method of performing optically-driven and laser-enhanced rotational cooling. We then present our design to optically pump the system to the single stretched hyperfine state. The simulation details are described in Section III while Section IV presents and discusses our results. We conclude in Section V.

\begin{figure*}[!htp]
\centering
\includegraphics[width=18cm]{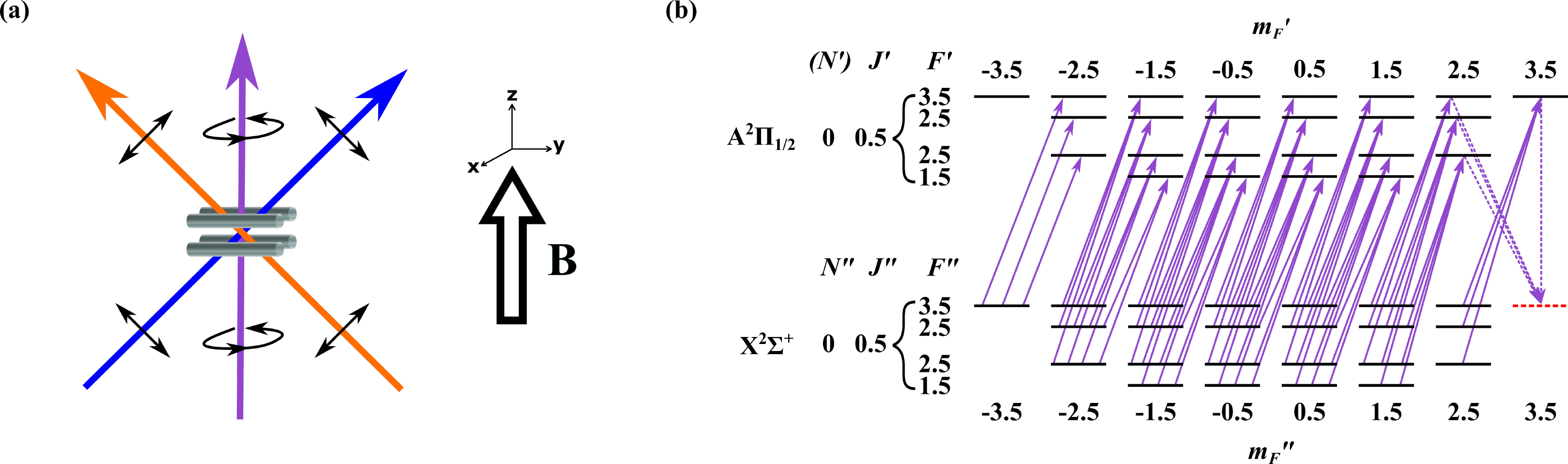}
\caption{(a) A schematic of the experimental setup and (b) the hyperfine structure of the rovibrational ground state of $\mathrm{X}^2\Sigma^+$ and $\mathrm{A}^2\Pi_{1/2}$. In (a), quadrupole rods of a linear Paul trap and two SFFLs are shown. The blue arrow represents the linearly-polarized SFFL (RC laser) that performs rotational cooling. The light-purple arrow represents the $\sigma^+$-polarized SFFL (HC laser) that drives the $\mathrm{Q}_{11}(0.5)$-branch transition, optically pumping the population into the stretched hyperfine state of the ground rovibronic manifold. The orange arrow represents the infrared laser (V10 laser) that drives the $1$\,--\;$0$ vibrational transition. In (b), the hyperfine structure and the set of transitions driven by the $\sigma^+$-polarized SFFL are shown. States of a given energy level are non-degenerate due to Zeeman shifts (not drawn; $\Delta E_Z \ll \Delta E_F$). The population in the rovibrational ground state of $\mathrm{X}^2\Sigma^+$ is driven towards a single dark state, $\lvert \mathrm{X}^2\Sigma^+,\, v=0,\, N=0,\, F''=3.5,\, m_{F''}=3.5\rangle$, represented by the dashed red line. The solid purple arrows represent transitions driven by the HC laser. The dashed purple arrows represent spontaneous emission channels to the target state.}
\label{schematic_setup}
\end{figure*}

\section{Theory and Methods}
In the electronic ground state of $\mathrm{AlH}^+$, $\mathrm{X}^2\Sigma^+$, the angular momenta are well-described by Hund's case (b), with good quantum numbers  $\left\lbrace\right.\! \Lambda, N,\, S,\, J,\, \pm \!\left.\right\rbrace$. $J\,(\vec{J}=\vec{N}+\vec{S}\;)$ is the quantum number of the total angular momentum exclusive of nuclear spin; $N$ is the rotational quantum number; and $S$ is the electron-spin quantum number. The $\mathrm{A}^2\Pi$ electronic state of $\mathrm{AlH}^+$ is better-described using the Hund's case (a) basis of $\left\lbrace\right.\! \Lambda,\, S,\, \Sigma,\, J, \Omega,\, \pm \!\left.\right\rbrace$. Here, $\Lambda$, $\Sigma$, and $\Omega$ are projections along the molecular axis of the electronic orbital angular momentum, electron-spin, and their sum ($\Lambda + \Sigma$), respectively. The eigenvalues of the parity operator, $P$, are represented as $\pm$.

Though $N$ is not a good quantum number in $\mathrm{A}^2\Pi$, for convenience, we still use $N$ to denote different $J$ values in the $\mathrm{A}^2\Pi$ state (see Figure \ref{level_diagram_parity}) and to label transition branches. It should be noted that the rotational states of both the $\mathrm{X}$ and $\mathrm{A}$ manifolds of $\mathrm{AlH}^+$ exhibit doublets. In the $\mathrm{X}$ manifold, the doubling is a result of the interaction of the electron-spin and molecular rotation whereas the doubling in the $\mathrm{A}^2\Pi$ manifold is produced by the coupling between electronic orbital angular momentum and molecular rotation (known as $\Lambda$-doubling). For both electronic states, $v$ is the vibrational quantum number. We invoke the convention that $v'$ ($v''$) denotes the vibrational quantum number of the upper (lower) state of a spectroscopic transition.

\subsection{Rotational Cooling}
Our group has previously demonstrated broadband rotational cooling of $\mathrm{AlH}^+$ using a linearly-polarized SFFL with 100 fs pulses centered at 360 nm. The pulses are generated from a frequency-doubled femtosecond laser (Spectra-Physics Mai Tai) with 80 MHz repetition rate, so the pulse spectrum is divided into a frequency comb with 80 MHz spacing between neighboring teeth. Its large bandwidth pumps many rotational transitions simultaneously. We use spectral filtering in the Fourier plane to selectively excite rotational cooling transitions. Within the $\mathrm{A}^2\Pi$\;--\;$\mathrm{X}^2\Sigma^+$ set of transitions, the $v'=1$\;--\;$v''=1$ band lies $\sim$ 150 cm$ ^{-1}$ above the 0\;--\;0 band. The neighboring 1\;--\;0 and 0\;--\;1 bands are also nearly a vibrational constant away from the 0\;--\;0 band$^{19}$. Thus, we ignored the 1\;--\;1 and 1\;--\;0 bands because they are outside the passband of our spectrally-filtered SFFLs. However, the 0\;--\;1 band was included because it provides a critical relaxation pathway. Informed by these choices, we only included the $\lvert \mathrm{X}^2\Sigma^+,\, v = 0,1 \rangle$ and $\lvert \mathrm{A}^2\Sigma^+,\, v = 0 \rangle$ states in our population dynamics model.

As shown in Figure \ref{level_diagram_parity}, the rotational cooling process has two parts. The first part is a fast cycle in which linearly-polarized 360 nm pulses of the SFFL drive the electronic transition connecting $\lvert \mathrm{X}^2\Sigma^+,\, v''=0\rangle$ and $\lvert \mathrm{A}^{2}\Pi_{1/2},\, v'=0\rangle$. Following excitation to the $\mathrm{A}$ state, electronic spontaneous emission without vibrational excitation occurs with a lifetime of $\sim$ 0.4 \si{\micro}s. The branching ratio for $\lvert \mathrm{A}^{2}\Pi_{1/2},\, v'=0\rangle$ $\rightarrow$ $\lvert \mathrm{X}^2\Sigma^+,\, v''=0\rangle$ is $\sim$ 0.97. Before rotational cooling, there are significant rotational populations in the lowest ten rotational states. By way of the fast cycle, nearly all of it can be driven into the two lowest rotational states, $\lvert v''=0,\, N''=0,1\rangle$, within a few microseconds. However, the parity is flipped for dipole transitions such as $\lvert \mathrm{A}^2\Pi\rangle$\;--\;$\lvert \mathrm{X}^2\Sigma^+\rangle$. As a result, the population in $\lvert \mathrm{X}^2\Sigma^+,\, v''=0,\, N''=1, -\rangle$ cannot be transferred to the rovibronic ground state, $\lvert \mathrm{X}^2\Sigma^+,\, v''=0,\, N''=0,\, +\rangle$ via this fast cycle because each cycle involves a pair of transitions that together conserve the parity. The population in $\lvert \mathrm{X}^2\Sigma^+,\, v''=0,\, N''=1, -\rangle$ can still transfer to $\lvert \mathrm{X}^2\Sigma^+,\, v''=0,\, N''=0,\, +\rangle$, but must do so in an odd number of transitions to flip the parity. The shortest parity-flipping process happens in three transitions: first, the population in $\lvert \mathrm{X}^2\Sigma^+,\, v''=0,\, N''=1, -\rangle$ is excited by the SFFL to the $\mathrm{A}$ state; then there is spontaneous decay to an intermediate state with negative parity, $\lvert \mathrm{X}^2\Sigma^+,\, v''=1,\, N''=1,\, -\rangle$; last, the population undergoes a vibrational relaxation to reach $\lvert \mathrm{X}^2\Sigma^+,\, v''=0,\, N''=0,\, +\rangle$. This parity-flipping process, which constitutes the second part of the cooling process, is relatively slow because the vibrational lifetime for the $\lvert \mathrm{X}^2\Sigma^+,\, v=1\rangle$\;--\;$\lvert \mathrm{X}^2\Sigma^+,\, v=0\rangle$ transition is 140 ms.

In our laboratory, $\mathrm{AlH}^+$ molecular ions are held in a linear Paul trap and sympathetically cooled to sub-Kelvin translational temperatures using co-trapped Doppler-cooled $\mathrm{Ba}^+$ atomic ions. To remove the dark states of the barium ion during the Doppler cooling process, a 2 G magnetic field is applied. After translational cooling, the linearly-polarized SFFL (with polarization of 45$^{\circ}$ relative to the direction of the magnetic field) is turned on to rotationally cool the molecules into their rovibronic ground state.\par

Rotational cooling of the negative-parity populations is rate-limited by the vibrational-decay timescale. We propose to address this bottleneck through the addition of a 6.7 \si{\micro}m continuous-wave laser (V10) which drives the $\lvert \mathrm{X}^2\Sigma^+,\, v''=1,\, N''=1,\, -\rangle$\;--\;$\lvert \mathrm{X}^2\Sigma^+,\, v''=0,\, N''=2,\, +\rangle$ transition to accelerate the parity-flipping process. This technique has not yet been applied to rotational cooling using broadband lasers. We simulate the rotational cooling process and show that it is accelerated by the additional laser. The simulation results are summarized in Figure \ref{RC_RCV10} and Table \ref{RC_RC_V10table}.

\subsection{Hyperfine Cooling}
$\mathrm{AlH}^+$ has one unpaired electron ($S=\frac{1}{2}$), and nuclei with nuclear spins $I_\mathrm{Al}=\frac{5}{2}$ and $I_\mathrm{H}=\frac{1}{2}$. We define the hyperfine states of the electronic ground state of $\mathrm{AlH}^+$ in terms of a set of total angular momentum quantum numbers, $\left\lbrace F \right\rbrace$. We define the members of the set as follows: $F_{1}=J+I_\mathrm{Al}$, and $F=F_{1}+I_\mathrm{H}$. In its rovibronic ground state ($\mathrm{X}$, $v=0$, $N=0$), $\mathrm{AlH}^+$ thus has four hyperfine states: $F=\left\lbrace\frac{3}{2},\, \frac{5}{2}\right\rbrace$ for $F_{1}=2$ and $F=\left\lbrace\frac{5}{2},\, \frac{7}{2}\right\rbrace$ for $F_{1}=3$.

As we are interested in pumping our system to a single hyperfine state, we also added a circularly-polarized 360 nm beam propagating along the direction of the 10 G magnetic field as shown in Figure \ref{schematic_setup}(a). By taking advantage of the selection rules for dipole transitions driven by $\sigma^+$-polarized light, we can optically pump the system to the stretched hyperfine state in which the total angular momentum ($F$) has the largest projection ($m_F$) along the quantization axis. The schematic plot of our setup is shown in Figure \ref{schematic_setup}(a). When we apply the $\sigma^+$-polarized laser, the selection rule are as follows:

\begin{align*}
\Delta F=0,\, \pm 1 \\
\Delta m_F=1
\end{align*}

As can be seen in Figure \ref{schematic_setup}(b), if we set the $\sigma^+$-polarized laser to drive the $\mathrm{Q}_{11}(0.5)$-branch transition\footnote[2]{This notation describes the branch in terms of both $N$ and $J$: $^{\Delta N}\Delta J_{ul}\left( J'' \right)$. Note that $N$ is not a good quantum number in $\mathrm{A}^2\Pi$, and simply serves as a convenient label. If $\Delta N = \Delta J$, then the notation uses one letter to mark the type of branch. $u$ and $l$ denote the spin orientations of the upper and lower states of a transition. In our case, the upper state could be either $\mathrm{A}^2\Pi_{1/2}$ or $\mathrm{A}^2\Pi_{3/2}$. These states, corresponding to $\lvert \Omega=\Lambda -\frac{1}{2} \rangle$ or $\lvert \Omega=\Lambda +\frac{1}{2} \rangle$, are denoted as $u=1$ and $u=2$, respectively. Analogously, the lower-state, part of the $\mathrm{X}^2 \Sigma^{+}$ manifold, has $S=1/2$. Our convention is that $l=1$ and $l=2$ represent $\lvert J=N+\frac{1}{2} \rangle$ and $\lvert J=N-\frac{1}{2} \rangle$, respectively.}, $\lvert \mathrm{A}^2\Pi_{1/2},\, v'=0,\, N'=0\rangle$\,--\;$\lvert \mathrm{X}^2\Sigma^+,\, v''=0,\, N''=0\rangle$, most of the population in the rovibronic ground state of $\mathrm{X}^2 \Sigma^{+}$ will be further optically pumped to the stretched state of maximal $F$. This stretched hyperfine state is a dark state; it cannot absorb any more $\sigma^+$-polarized photons because there is no higher $m_F$-state available within the $\lvert \mathrm{A}^2\Pi_{1/2},\, v=0\,, N=0\rangle$ manifold. Thus the population will accumulate in the stretched state over time. The addition of the 6.7 \si{\micro}m continuous-wave laser also accelerates this hyperfine cooling process since it is dependent upon the rate of rotational cooling. The simulation results are shown in Figure \ref{RCHC_RCHCV10} and Table \ref{RCHC_RCHCV10table}.

\section{Simulation Details}

\begin{table*}
\centering
\begin{minipage}[b]{0.45\linewidth}
\renewcommand*{\thempfootnote}{\fnsymbol{mpfootnote}}
\let\TPToverlap=\TPTrlap
\caption{Molecular constants for the $\mathrm{X}^2\Sigma^+$ state of $\mathrm{AlH}^+$\tnote{$\ddag$}}
\begin{threeparttable}
\renewcommand{\arraystretch}{1.25}
\setlength{\tabcolsep}{1em}
\begin{tabular}{lcc}
 \hline
Constant\cite{szajna2011high}     &    $ v=0$   &   $ v=1$    \\
 \hline
$B_{v}$                           &$6.563231$   &$6.184845$    \\   \hline
$D_{v}\times 10^{4}$              &$4.5720$     &$5.0983$      \\ \hline
$H_{v}\times 10^{8}$              &$-0.238$     &$-6.586$       \\ \hline
$L_{v}\times 10^{11}$             &$-1.712$     &               \\ \hline
$M_{v}\times 10^{14}$             &$1.140$      &               \\ \hline
$N_{v}\times 10^{18}$             &$-7.07$      &               \\ \hline
$\gamma_{v}\times 10^{2}$         &$5.665$      &$5.035$        \\ \hline
$\gamma_{Dv}\times 10^{5}$        &$-1.896$     &$-2.09$        \\ \hline
origin                                          &$0$              &$1523$        \\ 
\hline
\end{tabular}
\begin{tablenotes}
\item[$\ddag$] in $cm^{-1}$
\end{tablenotes}
\end{threeparttable}
\label{Xparameters}

\end{minipage}
\hspace{0.2cm}
\begin{minipage}[b]{0.45\linewidth}
\centering
\let\TPToverlap=\TPTrlap
\caption{Molecular constants for the $\mathrm{A}^2\Pi$ state of $\mathrm{AlH}^+$\tnote{$\ddag$}}
\renewcommand{\arraystretch}{1.25}
\begin{threeparttable}
\setlength{\tabcolsep}{1em}
\begin{tabular}{lc}
 \hline
Constant                                  &                $ v=0$   \\
 \hline
$B_{v}$                                   &$6.727$ \cite{almy1934band}     \\   \hline
$A_{v}$                                   &$108$ \cite{almy1934band}        \\ \hline
$p_{v}\times 10^{2}$             &$1.643$ \cite{szajna2011high}    \\ \hline
$q_{v}\times 10^{3}$             &$1.499$ \cite{szajna2011high}    \\ \hline
$D_{v}\times 10^{4}$             &$-4.14$ \cite{almy1934band}     \\ \hline
origin                                      &$27713$ \cite{szajna2011high}   \\ 
\hline
\end{tabular}
\begin{tablenotes}
\item[$\ddag$] in $cm^{-1}$
\end{tablenotes}
\end{threeparttable}
\label{Aparameters}
\end{minipage}
\end{table*}

\begin{center}
\begin{table*}
\let\TPToverlap=\TPTrlap
\centering
\small
  \caption{Permanent and transition dipole moments ($\langle$i$\lvert\,\hat{\mu}\,\rvert$j$\rangle$)\tnote{$\dag$} \tnote{$\ddag$}}
  \renewcommand{\arraystretch}{1.5}
\begin{threeparttable}
\setlength{\tabcolsep}{3pt}
\begin{tabular*}{0.48\textwidth}{@{\extracolsep{\fill}}cccc}
\hline
\diagbox{State i}{State j} &$\mathrm{X}^2\Sigma^+,\, v=0$ &$\mathrm{X}^2\Sigma^+,\, v=1$    &$\mathrm{A}^2\Pi,\, v=0$  \\ 
\hline
$\mathrm{X}^2\Sigma^+,\, v=0$         &$-0.389$               &                      & \\   \hline
$\mathrm{X}^2\Sigma^+,\, v=1$         &$0.087$                &$-0.2861$             & \\   \hline
$\mathrm{A}^2\Pi,\, v=0$              &$1.566$                &$-0.2806$             &$-0.928$\\ \hline
\end{tabular*}
\begin{tablenotes}
\item[$\dag$] The signs of the dipole moments reflect the choice of coordinate system in which the lighter atom was placed at $+\hat{z}$.
\item[$\ddag$] in debye
\end{tablenotes}
\end{threeparttable}
\label{dipolemoment}
\end{table*}
\end{center}

\begin{table*}[htbp!]
\centering
\let\TPToverlap=\TPTrlap
\caption{Hyperfine constants of $\mathrm{AlH}^+$\tnote{$\ddag$}}
\renewcommand{\arraystretch}{1.5}
\begin{threeparttable}
\setlength{\tabcolsep}{6pt}
\begin{tabular}{*{5}{c}}
\hline
\multirow{2}{*}{Constant} &\multicolumn{2}{c}{$\mathrm{X}^2\Sigma^+$} &\multicolumn{2}{c}{$A^2\Pi$} \\
\cline {2-5}
&Al &H  &Al &H\\
 \hline
a               &\num{1.680E-3}           &\num{8.64E-5}            &\num{8.511E-3}              &\num{5.53E-4}         \\
b               &\num{3.951E-2}           &\num{2.01E-2}             &\num{1.382E-2}             &\num{-9.35E-3}         \\
c               &\num{5.039E-3}          &\num{2.59E-4}             &\num{-5.654E-3}            &\num{2.79E-4}         \\
d               &\num{0}                        &\num{0}                         &\num{1.040E-2}             &\num{4.60E-4}          \\
$eQq_0$    &\num{-1.341211E-3}    &\num{2.12636E-6}        &\num{6.20358E-4}       &\num{2.34231E-6}      \\
$eQq_2$    &\num{0}                        &\num{0}                          &\num{-2.89652E-3}    &\num{-5.63315E-7}     \\               
\hline
\end{tabular}
\begin{tablenotes}
\item[$\ddag$] in cm$^{-1}$
\end{tablenotes}
\end{threeparttable}
\label{hyperfineconstant}
\end{table*}

Our population dynamics were modeled by the following system of rate equations:
\begin{align}
\begin{aligned}
\frac{\partial\rho_i}{\partial t}=-\sum_{j\neq i}B_{ij}(I_{\mathrm{BBR}}+I_{\mathrm{laser}})\rho_i - \sum_{j<i}A_{ij}\rho_i \\
+\sum_{j\neq i}B_{ji}(I_{\mathrm{BBR}}+I_{\mathrm{laser}})\rho_i + \sum_{j>i}A_{ji}\rho_i 
\end{aligned}
\end{align}
where $\rho_i$ is the population fraction in state $i$. The system of equations includes the rovibronic and hyperfine states of interest. The initial population was assumed to be thermal with a temperature of 300 K. $I_{\mathrm{BBR}}$ and $I_{\mathrm{laser}}$ are the energy densities of the blackbody radiation and laser. $A$ and $B$ are the spontaneous emission and stimulated emission Einstein coefficients, respectively. The Einstein coefficients can be expressed using the following equations:
\begin{align}
\begin{aligned}
A_{ul} &= \frac{2\pi \widetilde{\nu}^2 q_e^2}{\epsilon_0 m_e c} \frac{g_l}{g_u} f_{lu}\\
B_{ul} &= \frac{q_e^2}{4 \epsilon_0 m_e h c \widetilde{\nu}} \frac{g_l}{g_u} f_{lu} \\
B_{lu} &= \frac{q_e^2}{4 \epsilon_0 m_e h c \widetilde{\nu}} f_{lu}\\
\end{aligned}
\end{align}
where $\widetilde{\nu}$ is the wavenumber energy of the transition; $q_e$ and $m_e$ are the charge and the rest-mass of an electron; $\epsilon_0$ is the vacuum permittivity; c is the speed of light; $g_l$ and $g_u$ are the degeneracies of the lower and the upper states, respectively; and $f_{lu}$ is the oscillator strength of the transition.
In order to determine the Einstein coefficients using Equation (2), 
we utilized Western's \texttt{PGOPHER}\cite{western2017pgopher} software to 
compute transition energies and oscillator strengths for 
$\mathrm{AlH}^+$. In the absence of the magnetic field, for $J$\,--\;$J+1$ rovibronic transitions with linearly-polarized light, the two stretched states, $\lvert J\; m_J\rangle = \left\lbrace\lvert\left(J+1\right)\;\pm\left(J+1\right)\rangle\right\rbrace$, are inaccessible by electric dipole transitions and constitute dark states as shown in Figure \ref{level_diagram_parity}(b). A 10 G magnetic field was applied to address this problem while driving the P-branch rotational cooling transition with a linearly-polarized beam. By adding the 10 G magnetic field at an angle of 45$^{\circ}$ with respect to the polarization direction of the rotational cooling laser (see Figure \ref{schematic_setup}(a)), bright states are mixed with the dark state. The brightened state evolves at the Larmor frequency (10$^9$ s$^{-1}$), which is sufficiently fast compared to the Rabi frequency of the rotational cooling laser (10$^8$ s$^{-1}$) to destabilize the dark population and expose it to the cooling laser. \texttt{PGOPHER} required a number of empirical parameters to describe the states of $\mathrm{AlH}^+$. Table \ref{Xparameters} and \ref{Aparameters} present the values we used to describe the $\mathrm{X}^2 \Sigma^{+}$  and $\mathrm{A}^2\Pi$ states.

We used Le Roy's \texttt{LEVEL}\cite{le2017level} to calculate vibrationally-averaged permanent and transition electric-dipole moments from potential-energy, permanent and transition electric-dipole moment functions of a prior work\cite{nguyen2011challenges}. These results are presented in Table \ref{dipolemoment}.

\texttt{Dalton}\cite{daltonpaper, daltonwebpage} quantum computational 
software was used to compute the hyperfine and nuclear electric-quadrupolar 
coupling constants of the $\mathrm{X}^2\Sigma^+$ and 
$\mathrm{A}^2\Pi$ states at fixed geometry ($R = 
\SI{1.6018}{\angstrom}$). We chose the pcJ-1 basis 
set\cite{Jensen2010} because the pcJ-$n$ family was optimized for 
calculating indirect nuclear spin-spin coupling constants and has 
tight functions that are well-suited for describing the electron 
density near the nucleus\footnote[3]{We performed test computations at the same level of theory for $\mathrm{HCl}^+$ and $\mathrm{OH}$ and found that the calculated hyperfine coupling constants for these two molecules were within 10--15 \% of published experimental values. We feel the relative agreement justifies our choices.}. The computations invoked a restricted active space self-consistent field (RASSCF)\cite{Olsen1988} wave function with 2,439 determinants for the $\mathrm{X}^2\Sigma^+$ state and 1,947 determinants for the $\mathrm{A}^2\Pi$ state. The wave function was defined by 5 inactive orbitals in the RAS1 space, a full-valence (5-orbital) RAS2 space, and single and double excitations from the RAS2 space into the RAS3 space for the 30 remaining orbitals. \texttt{Dalton} outputs the values of the hyperfine tensor components ($A_{xx},\, A_{yy},\, A_{zz}$) and the Fermi contact term ($A_{iso}$). However, \texttt{PGOPHER} takes as inputs of hyperfine constants the Frosch--Foley 
coefficients ($a$, $b$, $c$, $d$). The conversion formulas are listed below:

\begin{align}
\label{eqn:spin-dipolar}
\begin{aligned}
  c &= \frac{3}{2} A_{zz} \\
  d &= A_{xx}-A_{yy} \\
  b &= A_{iso}-\frac{c}{3} \\
  a &= d + \frac{c}{3}
\end{aligned}
\end{align}

We used the electric-field gradients ($q_{xx}\equiv \partial ^{2}V_{x}/\partial x^{2}$, $q_{yy}\equiv \partial ^{2}V_{y}/\partial y^{2}$, $q_{zz}\equiv \partial ^{2}V_{z}/\partial z^{2}$ in $\mathrm{MHz}$) and the nuclear electric-quadrupole moment ($Q$ in barn) computed by \texttt{Dalton} to calculate the nuclear electric-quadrupolar coupling constants, $eQq_0$ and $eQq_2$ (in $\mathrm{cm^{-1}}$). The formulas are given by:

\begin{align}
\label{eqn:nuclear-quadrupole}
\begin{aligned}
  eQq_0 &= 7.8375814 \times 10^{-3} \mathrm{cm^{-1}} (\frac{q_{zz}}{\mathrm{MHz}})(\frac{Q}{\mathrm{barn}})\\
  eQq_2 &= 7.8375814 \times 10^{-3} \mathrm{cm^{-1}} (\frac{q_{xx}-q_{yy}}{\mathrm{MHz}})(\frac{Q}{\mathrm{barn}})\\
\end{aligned}
\end{align}

The input hyperfine coupling constants for \texttt{PGOPHER} are listed in Table \ref{hyperfineconstant}.

At room temperature, 99$\, \%$ of the $\mathrm{AlH}^+$ population is in the lowest vibrational state, $v=0$, within the $\mathrm{X}^2 \Sigma^+$ manifold. In turn, within this vibrational ground state, thermal distribution produces significant populations among the first ten $J$-levels, $J=\left\lbrace 0.5,\hdots, 9.5 \right\rbrace$, and less than 4$\, \%$ in $J>9.5$. Therefore, we included the lowest ten $J$-states ($J=\{0.5,...,9.5\}$) of each vibrational state (i.e. $\lvert \mathrm{X}^2\Sigma^+,\, v = 0,\,1 \rangle$ and $\lvert \mathrm{A}^2\Pi,\, v = 0 \rangle$) in the rate equation simulation. This manifold of states was able to resolve the vibronic relaxation from $\lvert \mathrm{A}^2\Pi,\, v = 0 \rangle$ to $\lvert \mathrm{X}^2\Sigma^+,\, v = 1 \rangle$ and the parity-flipping process via the intermediate states.

Our femtosecond laser was given a frequency-domain representation in the simulation. The spectrum was described by 80 MHz-spaced comb teeth within a Gaussian envelope of $\sim$ 2.6 nm FWHM bandwidth and centered at 360 nm. We modeled our spectral filtering apparatus as a cut-off filter to the laser spectrum. The cut-off frequency was chosen so as to pass the range of frequencies that selectively drove a set of cooling transitions. For example, to rotationally cool we chose to drive the $\mathrm{P}_{11}$, $\mathrm{^OP}_{12}$ and the $\mathrm{^PQ}_{12}$ branches using the linearly-polarized SFFL.  To subsequently cool the hyperfine manifold, we added the $\mathrm{Q}_{11}(0.5)$ branch and drove it with the $\sigma^+$-polarized SFFL. Informed by our previous experimental work, we split the 200 mW laser power equally between the linear- and the $\sigma^+$-polarized beams, each having a focused 400 \si{\micro}m-diameter spot at the center of the ensemble of $\mathrm{AlH}^+$ molecules.

The typical optical transition linewidth for $\mathrm{AlH}^+$ is $\sim$ 20 MHz, which is smaller than the 80 MHz comb-teeth spacing of the femtosecond laser spectrum. As a result, it was possible for an SFFL comb-line to fall outside of the transition linewidth of some of the transitions we desired to drive. We solved this issue in the simulation by introducing a Doppler-broadened linewidth contribution corresponding to a $\sim$ 1 K translational temperature for the $\mathrm{AlH}^+$ ion cloud. Doing so ensured that there was at least one comb-line within the linewidth of every desired cooling transition. One can accomplish this form of broadening in the experiment by raising the translational temperature of the $\mathrm{AlH}^+$ ions in a couple ways. One can excite the secular motion of the $\mathrm{AlH}^+$ with an AC field or introduce additional micro-motion by shifting the entire ion cloud away from the geometric center of the Paul trap using a DC field. After internal cooling is finished, one can then turn off the source of translational heating, allowing the $\mathrm{AlH}^+$ to be sympathetic cooled once again by the laser-cooled $\mathrm{Ba}^+$ atoms.

\begin{figure*}[htbp!]
\centering
\includegraphics[width=12cm]{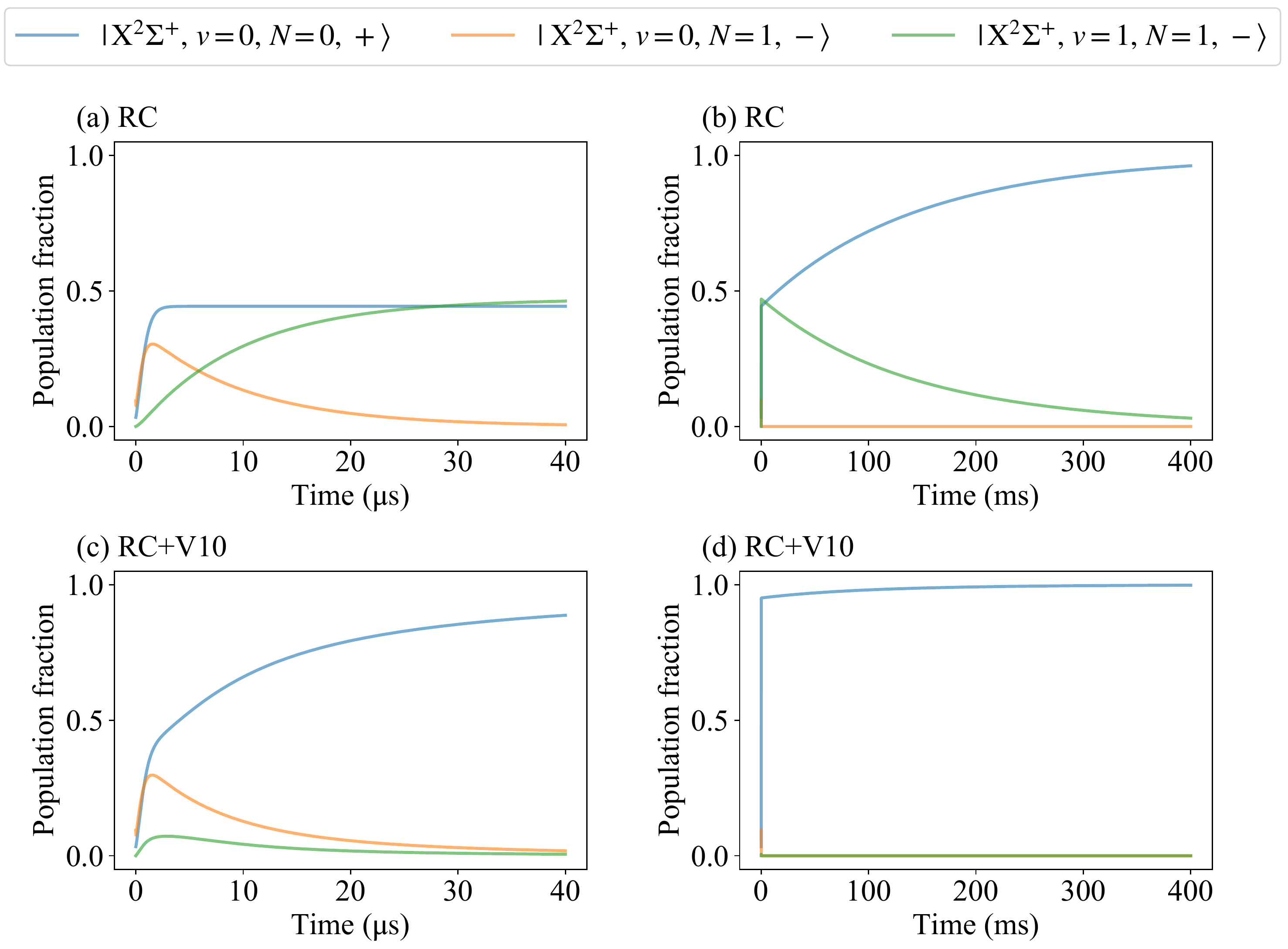}
\caption
{Simulated population dynamics of the rotational cooling: The plots in the top row show rotational cooling performed by application of the linearly-polarized SFFL (RC) for (a) 40 \si{\micro}s and (b) 400 ms, respectively. The plots in the bottom row show rotational cooling performed by RC while enhanced by the V10 laser for (c) 40 \si{\micro}s and (d) 400 ms, respectively. The V10 laser drives the $\lvert \mathrm{X}^2\Sigma^+,\, v''=1,\, N''=1,\; - \;\rangle$\,--\;$\lvert \mathrm{X}^2\Sigma^+,\, v''=0,\, N''=2,\, +\rangle$ transition.
}\label{RC_RCV10}
\end{figure*}

\begin{figure*}[htbp!]
\centering
\includegraphics[width=18cm]{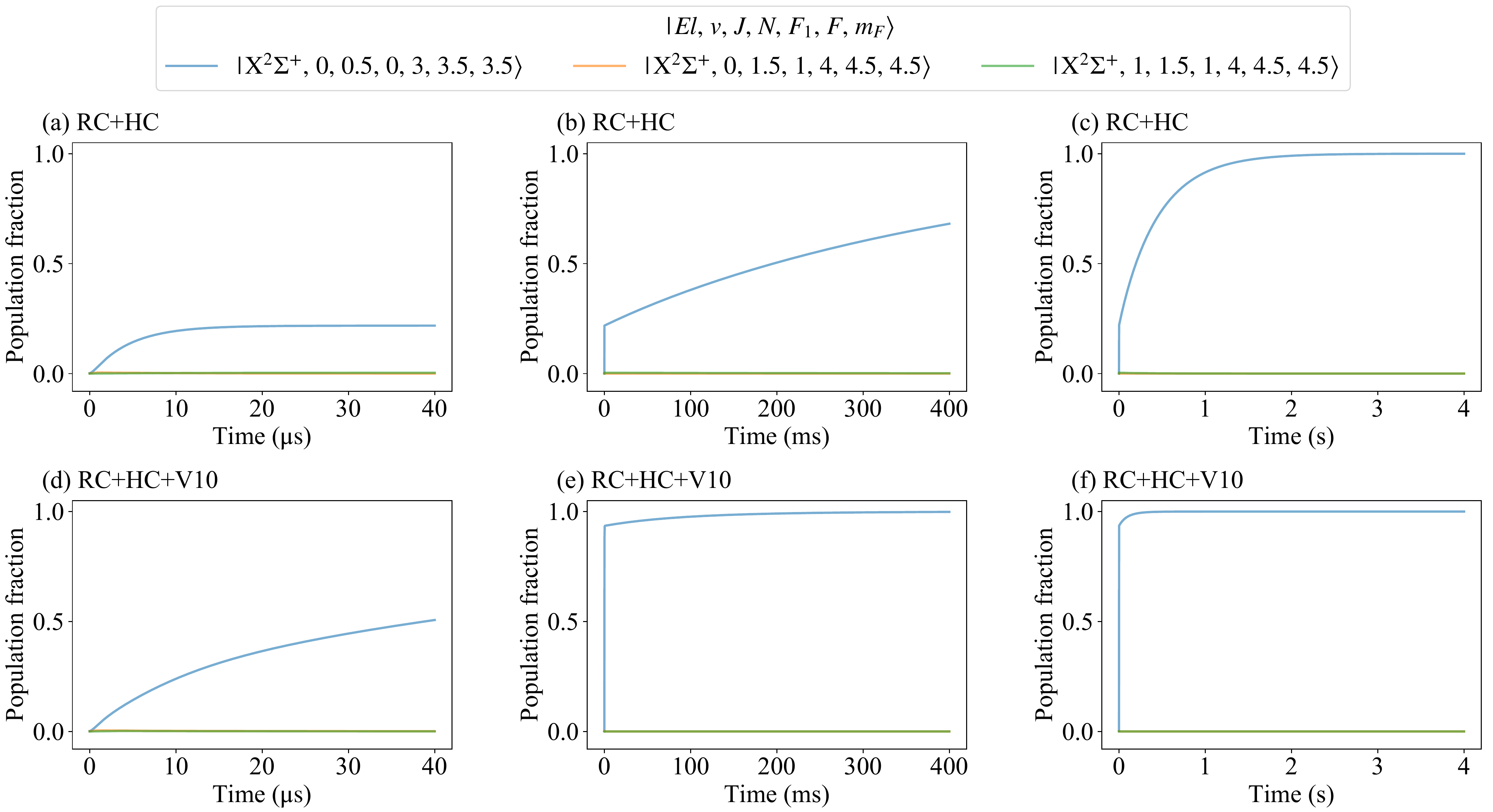}
\caption
{Simulated population dynamics of the single hyperfine-state preparation: The plots in the top row show hyperfine cooling performed by application of the linearly-polarized SFFL (RC) and the $\sigma^+$-polarized SFFL (HC) for (a) 40 \si{\micro}s, (b) 400 ms, and (c) 4 s, respectively. The plots in the bottom row show hyperfine cooling performed by RC and HC while enhanced by the V10 laser for (d) 40 \si{\micro}s, (e) 400 ms, and (f) 4 s, respectively. The V10 laser drives the $\lvert \mathrm{X}^2\Sigma^+,\, v''=1,\, N''=1,\, -\rangle$\,--\;$\lvert \mathrm{X}^2\Sigma^+,\, v''=0,\, N''=2,\, +\rangle$ transition.
}\label{RCHC_RCHCV10}
\end{figure*}

\begin{table*}[!htbp]
\begin{minipage}[b]{0.45\linewidth}
\centering
\renewcommand{\arraystretch}{1.25}
\caption{Population rise times for the rovibronic ground state}
\setlength{\tabcolsep}{6pt}
\begin{tabular}{ccc}
\hline
Laser fields & $63$ $\%$ &  $95$ $\%$   \\ 
\hline
RC & $60$ ms & $360$ ms \\ \hline
RC, V10 & $8.7$ \si{\micro}s & $160$ \si{\micro}s \\ 
\hline
\end{tabular}
\label{RC_RC_V10table}

\end{minipage}
\hspace{0.2cm}
\begin{minipage}[b]{0.45\linewidth}
\centering
\renewcommand{\arraystretch}{1.25}
\caption{Population rise times for $\lvert F=\frac{7}{2},\,m_F=\frac{7}{2}\rangle$ in the rovibronic ground state}
\setlength{\tabcolsep}{6pt}
\begin{tabular}{ccc}
\hline
Laser fields & $63$ $\%$ &  $95$ $\%$   \\ 
\hline
RC, HC &$330$ ms & $1200$ ms \\ \hline
RC, HC, V10 & $68$ \si{\micro}s & $25$ ms \\
\hline
\end{tabular}
\label{RCHC_RCHCV10table}
\end{minipage}
\end{table*}

We simulated the laser-enhanced parity-flipping process as well. For such cases, we represented the infrared laser (V10) that drives the $\lvert \mathrm{X}^2\Sigma^+,\, v''=1,\, N''=1,\, -\rangle$\,--\;$\lvert \mathrm{X}^2\Sigma^+,\, v''=0,\, N''=2,\, +\rangle$ transition to match the specifications of a commercial Fabry-Perot quantum-cascade laser from Thorlabs. This laser outputs $\sim$ 200 mW with a bandwidth of $\sim$15 cm$^{-1}$ (FWHM) and can be tuned to lase $\sim$ 6.7 \si{\micro}m, making it capable of driving the $v'=1$\,--\;$0$ line in the electronic ground state of $\mathrm{AlH}^+$.\\\\

\section{Results and discussion}
Figure \ref{RC_RCV10} and Table \ref{RC_RC_V10table} present rotational cooling rates for two schemes. In the first scheme, we apply the linearly-polarized rotational cooling laser (RC). In the second scheme, we apply the rotational cooling laser (RC) as well as the infrared laser (V10) that drives the $\lvert \mathrm{X}^2\Sigma^+,\, v''=1,\, N''=1,\, -\rangle$\,--\;$\lvert \mathrm{X}^2\Sigma^+,\, v''=0,\, N''=2,\, +\rangle$ transition. From Figure \ref{RC_RCV10}(a), it can be seen that without the V10 laser, the population in the rovibrational ground state ($v$,\,$N$) = (0,0) increases to 45 $\%$ within a few microseconds through the fast rotational cooling cycle. Afterwards, the population in (0,0) continues to increase but with a slower rate as shown in Figure \ref{RC_RCV10}(b). This behavior can be attributed to two time scales. At shorter times, population accumulates in the $\lvert \mathrm{X}^2\Sigma^+,\, v''=0,\, N''=1,\, -\rangle$ state after it undergoes a parity flip via the $\lvert \mathrm{X}^2\Sigma^+,\, v''=1,\, N''=1,\, -\rangle$ state. At longer times, vibrational relaxation (140 ms decay constant) begins to dominate the process. The addition of the V10 laser mitigates the effect of vibrational relaxation between $\lvert \mathrm{X}^2\Sigma^+,\, v''=1,\, N''=1,\, -\rangle$ and $\lvert \mathrm{X}^2\Sigma^+,\, v''=0,\, N''=2,\, +\rangle$. The time it takes for the population in the rovibronic ground state, $\rho_{0}^\mathrm{R}$, to grows to 63 $\%$ reduces from 60 ms to 8.7 \si{\micro}s. The trend continues as $\rho_{0}^\mathrm{R}$ reaches 95 $\%$ in only 160 \si{\micro}s, also significantly shorter than the 360 ms required in the absence of the V10 laser. 
However, the V10 laser addresses most but not all of the populations that accumulate in the $\lvert \mathrm{X}^2\Sigma^+,\, v'' = 1\rangle$ manifold. The populations in the higher rotational states of $\lvert \mathrm{X}^2\Sigma^+,\, v'' = 1\rangle$ either relax and re-enter the cooling cycle or undertake sequences of rotational and/or vibrational relaxations to reach the rovibronic ground state directly. Thus, as can be seen in Figure 3(d), after reaching $\sim$ 95\,\%, the growth rate of the rovibronic ground state population slows and becomes asymptotic.

Figure \ref{RCHC_RCHCV10} and Table \ref{RCHC_RCHCV10table} present our simulation results for two hyperfine-cooling schemes. In the first scheme, we apply the rotational cooling laser (RC) and the $\sigma^+$-hyperfine-cooling laser (HC). The second scheme adds the infrared laser (V10). As can be seen in Figure \ref{RCHC_RCHCV10}(a), in the absence of the V10 laser, the population in the stretched hyperfine state increases to $\sim 20 \%$ during the first tens of microseconds via the fast rotational cooling cycle and the hyperfine optical pump. Since the hyperfine cooling process takes additional cycles to transport the population to the stretched state, the time scale is longer when compared to the case during which only the RC laser is applied. The transfer rate eventually slows due to the relatively long vibrational relaxation time scale. After one second, the hyperfine population reaches more than 90 $\%$. From Figure \ref{RCHC_RCHCV10}(d), we can see that when the V10 laser is added, the impact of vibrational relaxation is reduced. The time it takes for the stretched hyperfine-state population in the rovibronic ground state, $\rho_{0}^{\mathrm{H}}$, to reach 63 $\%$ is shortened from 330 ms to 67 \si{\micro}s. If we leave the lasers on, $\rho_{0}^{\mathrm{H}}$ can reach 95 $\%$ in 25 ms, a near 50-fold reduction from the 1200 ms duration in the absence of the V10 laser. At longer times, the growth rate of $\rho_{0}^{\mathrm{H}}$ slows down due to the more complicated relaxation dynamics of the higher rotational states in $\lvert \mathrm{X}^2\Sigma^+,\, v'' = 1\rangle$.

A final point that deserves consideration is the effect of the polarization purity---that is, just how well one prepares the polarization of each laser field. While we do not expect the polarization to have significant effect on the timescale of the cooling dynamics, it will set an asymptotic limit on the population transferred to the target state.

\section{Conclusions}
In order for the fields of molecular quantum computing and simulations to mature, simple, efficient, and precise qubit state preparation will be critical. We have described an improvement to our previous work, in which the previously rate-limiting step of a parity-flip is sped up by the addition of a new infrared laser. We further described an extension to our rotational cooling setup that should enable optical pumping to a single hyperfine state. In support of this work, it was necessary to compute hyperfine matrix terms for the $\mathrm{A}^2\Pi$ state of $\mathrm{AlH}^+$. Simulations show that we should be able to drive 95 $\%$ of an ensemble of $\mathrm{AlH}^+$ molecules to a single quantum state within 25 ms.

\section*{Conflicts of interest}
There are no conflicts to declare.

\section*{Acknowledgements}
The researchers gratefully acknowledge support for this work from AFOSR grant number FA9550-17-1-0352.
A.G.S.O.-F thanks the S\~ao Paulo Research Foundation (FAPESP) for grant 2020/08553-2, 
the Brazilian National Research Council (CNPq) for grant 306830/2018-3, and
the Coordena\'c\~ao de Aperfei\'coamento de Pessoal de N\'ivel Superior - Brasil (CAPES) - Finance Code 001.



\bibliography{main.bib} 
\bibliographystyle{ieeetr} 

\end{document}